\documentstyle[12pt,aaspp4]{article}

\lefthead{Witt, Friedmann, \& Sasseen}
\righthead{Far-UV Scattering Properties of Galactic Dust}

\begin{document}

\title{Radiative Transfer Analysis of Far-UV Background Observations 
	Obtained with the Far-Ultraviolet Space Telescope (FAUST)}

\author{Adolf N. Witt, Brian C. Friedmann}
\affil{Ritter Astrophysical Research Center, University of Toledo, Toledo, OH 43606}
\authoremail{anw@physics.utoledo.edu,bfriedm@physics.utoledo.edu}

\and

\author{Timothy P. Sasseen}
\affil{Center for EUV Astrophysics, \\
2150 Kittredge Street, University of California, Berkeley, CA 94720}
\authoremail{tims@ssl.berkeley.edu}

\begin{abstract}

In 1992 the Far-Ultraviolet Space Telescope (FAUST) provided measurements of the ultraviolet
(140-180nm) diffuse sky background at high, medium, and low Galactic latitudes.  A significant
fraction of the detected radiation was found to be of Galactic origin, resulting from scattering
by dust in the diffuse interstellar medium.  To
simulate the radiative transfer in the Galaxy, we employed a Monte Carlo model 
which utilized a realistic, non-isotropic 
radiation
field based on the measured fluxes (at 156nm) and positions of 58,000 TD-1 stars, and a cloud
structure for the interstellar medium.  The comparison of the model predictions with the
observations led to a separation of the Galactic scattered radiation from an approximately
constant background, attributed to airglow and extragalactic radiation, and to a well
constrained determination of the dust scattering properties.  The derived dust albedo $a~=~0.45
~\pm~0.05$ is substantially lower than albedos derived for dust in dense reflection nebulae and
star-forming regions, while the phase function asymmetry $g~=~0.68~\pm~0.10$ is indicative of a
strongly forward directed phase function.  We show the highly non-isotropic phase function to
be responsible, in conjunction with the non-isotropic UV radiation field, for the wide range of
observed correlations between the diffusely scattered Galactic radiation and the column
densities
of neutral atomic hydrogen.  The low dust albedo is attributed to a size distribution of grains
in the diffuse medium with average sizes smaller than those in dense reflection nebulae.

\end{abstract}


\section{Introduction}

     When illuminated by the galactic interstellar radiation field, dust in the interstellar
medium gives rise to scattered radiation known as diffuse galactic light (DGL).  Efforts to
observe the DGL at different wavelengths and attempts to derive scattering properties of the 
dust
grains from such data have extended over the past half-century, and corresponding work directed
at the far-ultraviolet spectral region has been ongoing during the past quarter-century.  A
detailed review of DGL studies was given by Witt (\cite{Witt90}), while Bowyer (\cite{B91}) and Henry
(\cite{Hen91}) reviewed in-depth the particular complications involved in the observation and
interpretation of the far-ultraviolet background radiation due to several sources, including the
DGL.

     Major challenges facing DGL studies in the far-ultraviolet include the need for
observations with high diffuse-source sensitivity which provide extensive sky coverage while
simultaneously offering the means to separate the flux from discrete sources, such as stars and
galaxies, from that of the diffuse background.  For the interpretation of DGL data, there has
been a demand (\cite{Mat93}) for multiple-scattering models which take into account the
inhomogeneity of the scattering interstellar medium and the anisotropic distribution of the
interstellar radiation field, which is extreme for the far-ultraviolet sky (\cite{MH95}). 
The successful flight of the Far-Ultraviolet Space Telescope (FAUST) (\cite{BSLW93}),
conducted as part of NASA's 1992 ATLAS-1 shuttle mission (\cite{CT88}), has been
a response to the first of these two challenges.  The photon-counting imaging detector of FAUST
(\cite{Lamp90}) allowed a separation of the diffuse background from that due to stars in
the field to a level where unresolved stars contribute less than one percent of the total 
diffuse
intensity in the field (\cite{CSB94}).  At the same time, this instrument provided 
sensitive
measurements of the diffuse background in fields covering over 1000 square degrees in
representative directions ranging from high to intermediate latitudes (\cite{SLBW95}). 
Sasseen et al. (1995) showed that the dominant component in the measured far-ultraviolet
background is due to starlight scattered by galactic dust by demonstrating a clear relationship
between spatial power spectra of the IRAS 100$\mu$m cirrus and the FAUST far-ultraviolet
diffuse background images.

     An initial attempt to model the FAUST ultraviolet background and to constrain the
scattering properties of the diffusely distributed dust was undertaken by Sasseen \& Deharveng
(\cite{SD96}).  This attempt, which used the simple model of Jura (\cite{Jura79}) for high-latitude dust
illuminated by a constant plane source, met with only limited success however.  The model was
unable to correctly reflect the anisotropy and the longitudinal variation of the intensity of 
the
illuminating radiation field.

     In this paper, we are using a substantially more sophisticated model to treat the radiative
transfer for the diffuse background regions observed by FAUST.  This model was first
introduced and described briefly by Witt \& Petersohn (\cite{WP94}, hereafter WP) in the context of an
analysis of ultraviolet background observations made by the Dynamics Explorer 1 spacecraft
(\cite{FCF89}).

     The organization of this paper is as follows.  In \S 2 we will summarize the data reduction
and present an initial analysis. This will establish that the measured intensities are a
combination of DGL, which is dependent on the column density of the dust in the line of sight,
and of roughly constant contributions from airglow and possible extragalactic diffuse background
radiation.  In \S 3 we will present the WP model in greater detail then was done in WP (\cite{WP94}). 
Following this, we carry out the radiative transfer analysis of the FAUST data in \S 4, followed
by a discussion of the implications of the derived dust properties for the likely processing of
interstellar dust in the diffuse medium away from the galactic plane in \S 5.  In \S 6 we will
summarize our results and formulate our conclusions.

\section{The Data}

     The data used for this analysis were obtained with the imaging telescope, FAUST, during a
March 1992 shuttle flight.  FAUST has a bandpass covering the range 140 - 180nm and a 7.6$\arcdeg$
circular field of view.  The detector was a photon-counting microchannel plate with a wedge-and-strip
position sensitive anode (\cite{LSBB86}).  The in-flight angular resolution of the all-reflecting 
camera (D=161mm, f/1.12) was approximately 3.8$\arcmin$, which permitted the identification and
removal of point sources down to a detection limit of 1~x~10$^{-14}$ erg s$^{-1}$ cm$^{-2}$ \AA$^{-1}$.
Each pixel found to contain excess flux due to a point source was replaced with a weighted average 
based on distance, of those surrounding pixels containing negligible flux from the source 
(Sasseen et al.\ 1995).

     A summary of the data used for this analysis is given in Table 1. 
\begin{deluxetable}{lccccrrr}
\tablecaption{Data Used in this Study \label{table1}}
\tablehead{
\colhead{FAUST Field} & \colhead{\#} & \colhead{Points} & \colhead{$\ell$} & \colhead{$b$} &
\colhead{$<$I(UV)$>$} & \colhead{$<$I(100$\mu$m)$>$} & \colhead{$<$N(HI)$>$}  \\
\colhead{} & \colhead{} & \colhead{} & \colhead{($\arcdeg$)} & \colhead{($\arcdeg$)} & 
\colhead{(PU)} & \colhead{(MJy/sr)} & \colhead{(10$^{18}$/cm$^{2}$)}  \\
\colhead{(1)} & \colhead{(2)} & \colhead{(3)} & \colhead{(4)} & \colhead{(5)} & 
\colhead{(6)} & \colhead{(7)} & \colhead{(8)} 
}
\startdata
Dorado & 2 & 49 & 266.2 & -44.4 & 1705$\pm$\phn 90 & 0.272$\pm$0.227 & 200.8$\pm$\phn 55.1 \nl
NGPKM & 8 & 64 & 250.1 & 76.3 & 900$\pm$\phn 56 & 2.055$\pm$0.435 & 261.5$\pm$\phn 17.5 \nl
M87 & 9 & 64 & 285.5 & 75.0 & 1431$\pm$\phn 75 & 2.389$\pm$0.367 & 242.8$\pm$\phn 17.8 \nl
Virgo P2 & 10 & 58 & 284.3 & 76.4 & 1023$\pm$\phn 51 & 2.255$\pm$0.445 & 245.5$\pm$\phn 18.0 \nl
Virgo P1 & 11 & 62 & 283.8 & 71.6 & 1052$\pm$137 & 2.315$\pm$0.241 & 191.6$\pm$\phn 29.8 \nl
Centaurus & 14 & 59 & 302.8 & 21.7 & 3169$\pm$174 & 8.624$\pm$1.104 & 747.1$\pm$\phn 76.0 \nl
M83 & 16 & 64 & 314.7 & 32.8 & 1777$\pm$147 & 4.878$\pm$0.536 & 446.0$\pm$\phn 52.4 \nl
NGC 6752 & 18 & 62 & 336.8 & -25.2 & 2190$\pm$289 & 4.263$\pm$0.862 & 539.0$\pm$\phn 50.3 \nl
Hydra 20a & 20 & 62 & 261.4 & 37.6 & 1376$\pm$221 & 3.470$\pm$0.531 & 539.8$\pm$\phn 69.3 \nl
Hydra 20b & 20 & 49 & 267.5 & 38.1 & 1211$\pm$184 & 3.054$\pm$0.619 & 478.5$\pm$\phn 37.1 \nl
Hydra 20c & 20 & 64 & 273.5 & 38.5 & 1361$\pm$228 & 2.550$\pm$0.610 & 450.2$\pm$\phn 36.6 \nl
Hydra 21a & 21 & 63 & 314.5 & 35.3 & 2200$\pm$227 & 5.409$\pm$0.673 & 539.0$\pm$\phn 63.2 \nl
Hydra 21b & 21 & 50 & 309.5 & 38.3 & 2379$\pm$241 & 8.165$\pm$1.209 & 774.7$\pm$\phn 72.5 \nl
Hydra 21c & 21 & 62 & 304.3 & 40.7 & 2179$\pm$210 & 7.063$\pm$1.152 & 709.7$\pm$103.3 \nl
\enddata
\end{deluxetable}
The list contains the same
FAUST fields as studied by Sasseen \& Deharveng (\cite{SD96}), to which we have added data from
the FAUST image \# 2 in Dorado.  The columns in Table 1 give, respectively, (1) the name of
the FAUST field (Sasseen \& Deharveng 1996); (2) the FAUST image number 
(\cite{BSLW93}); (3) the number of 0.5$\arcdeg$ x 0.5$\arcdeg$ pixels in 
each image for which the diffuse UV intensity was
determined; (4) \& (5) the galactic coordinates for the image center, (6) the average intensity 
of
the diffuse radiation in the 140-180nm FAUST band in units of photons cm$^{-2}$ s$^{-1}$ \AA$^{-1}$ sr$^{-1}$,
(hereafter referred to as "units"); (7) the average intensity of the diffuse IRAS background
radiation at 100$\mu$m (\cite{IPAC91}); and (8) the average HI column density taken from the Bell Lab
HI Survey (\cite{SGW92}), supplemented with data from the Parkes HI survey (\cite{CHH79}; 
\cite{HC79}) for declination $\delta < -40\arcdeg$.  The averages listed in columns (7) and
(8) were performed over the same 0.5$\arcdeg$ x 0.5$\arcdeg$ pixels for which the average UV intensities were
determined.

     The data reduction, point source removal, and the treatment of foreground light sources
for the FAUST images was described in detail by Sasseen et al. (1995).  An additional
discussion of the impact of the space shuttle environment on the astronomical observations and
the removal of such effects is provided by Lampton et al. (\cite{LSWB93}) and Chakrabarti et al. (\cite{CSLB93}). 
With contributions due to direct starlight effectively removed (\cite{CSB94}), the residual
UV intensities listed in column (6) of Table 1 should therefore contain mainly a sum of DGL,
residual airglow, and an essentially isotropic cosmic background.  

     Small, additional contributions to the diffuse emissions
from the galaxy are likely in the form of H$_{2}$
fluorescence (\cite{MHB90}) and of line emission from C IV at 155.0nm,
Si IV at 139.7nm and O [III] at 166.3nm (\cite{MB90}).  Martin et al. (1990)
detected about 75-150 units at low and intermediate latitudes for H$_{2}$ fluorescence, while Martin
\& Bowyer found $<$ 20 units for the listed line emissions.  These intensities are small compared
to the variations in the airglow component, and lacking the facility to separate them
spectroscopically in the FAUST experiment, we will include these emissions in the airglow
component in subsequent sections and use DGL to refer only to dust-scattered radiation.

     We confirm the presence of a strong DGL component in the data by plotting the observed
average UV intensities for our 14 target fields against the corresponding HI column densities in
Figure 1.  
\begin{figure}[htb]
\plotone{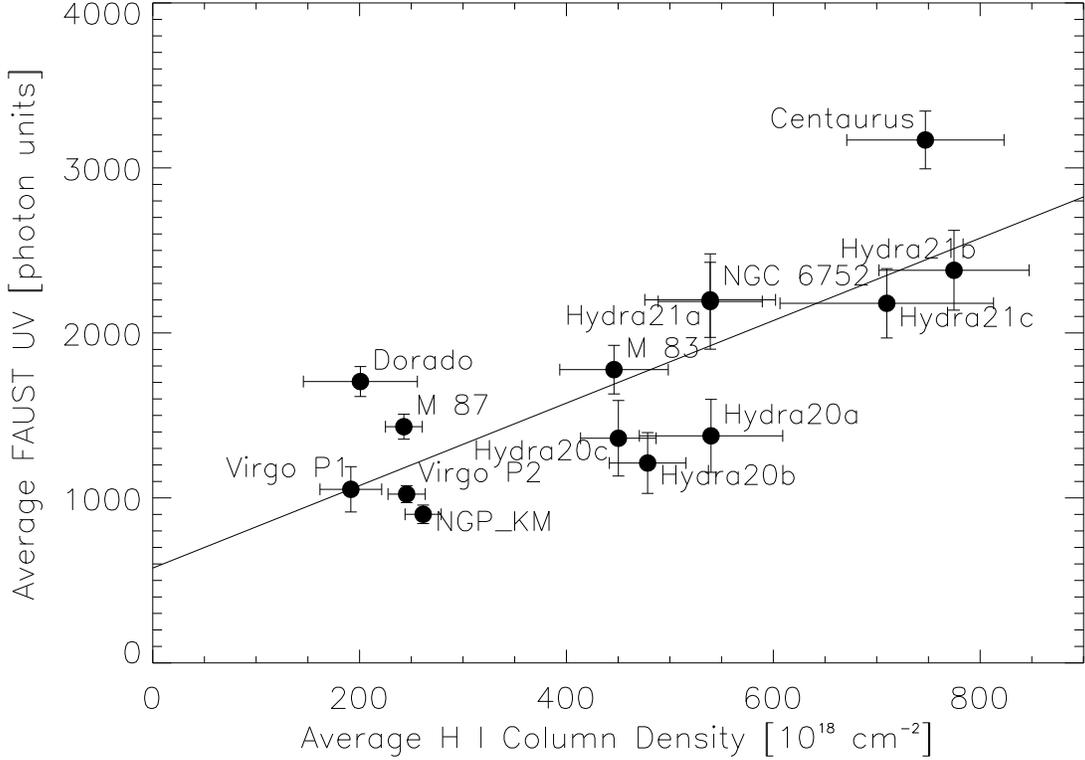}
\caption[f1.eps]{A plot of the average diffuse UV background intensity measured
in 14 FAUST fields against the average HI column densities.  Averages are performed
over a 4$\arcdeg$ x 4$\arcdeg$ area centered on the field coordinates.  The error bars 
represent one standard deviation of the distribution of values about the respective
averages. \label{fig1}}
\end{figure}
The DGL component is expected to depend upon the column density of dust and upon
the illumination conditions of this dust set by the galactic radiation field.  The column 
density
of dust is correlated with the HI column density (\cite{BSD78}).  Hence, we
expect a correlation between the DGL component and HI which Figure 1 indeed reveals.  

     We also note the interesting fact that the variation in the UV intensity in individual fields (Table 1)
compared to the variation of the HI column densities for the same fields is larger on average (ratio~=~ 
3.62~$\pm$~1.37 [photon units/10$^{18}$ cm$^{-2}$]) than the same ratio (Fig.1) for the 
total data set (2.50~$\pm$~0.37 [photon units/10$^{18}$ cm$^{-2}$]).  This is easily understood if
one considers the difference in spatial resolution represented by the two data sets, 30$\arcmin$ x 30$\arcmin$
pixels for the UV intensity measurements vs.\ the 2$\arcdeg$ FWHM width of the horn beam of the Bell
Lab HI survey.  We infer that there must be real intensity variations in the UV background
radiation due to structure in the ISM too small to be resolved by the HI survey.

     The positive intercept at N(HI) = 0 in Figure 1 suggests that the sum of residual airglow
and extragalactic background is approximately 600 units.  It is likely that this is an
underestimate because the coupling of a forward-throwing phase function with the 
concentration of galactic light sources near the galactic plane produces a higher scattered light 
intensity
per unit dust column at lower latitudes, i.e., higher N(HI) values, compared to higher 
latitudes,
or lower N(HI) values, thus steepening the slope of the correlation and lowering the
corresponding intercept (WP \cite{WP94}).  We can test this by plotting the UV
intensities against corresponding 100$\mu$m IRAS background intensities, as shown in Figure 2. 
\begin{figure}[htb]
\plotone{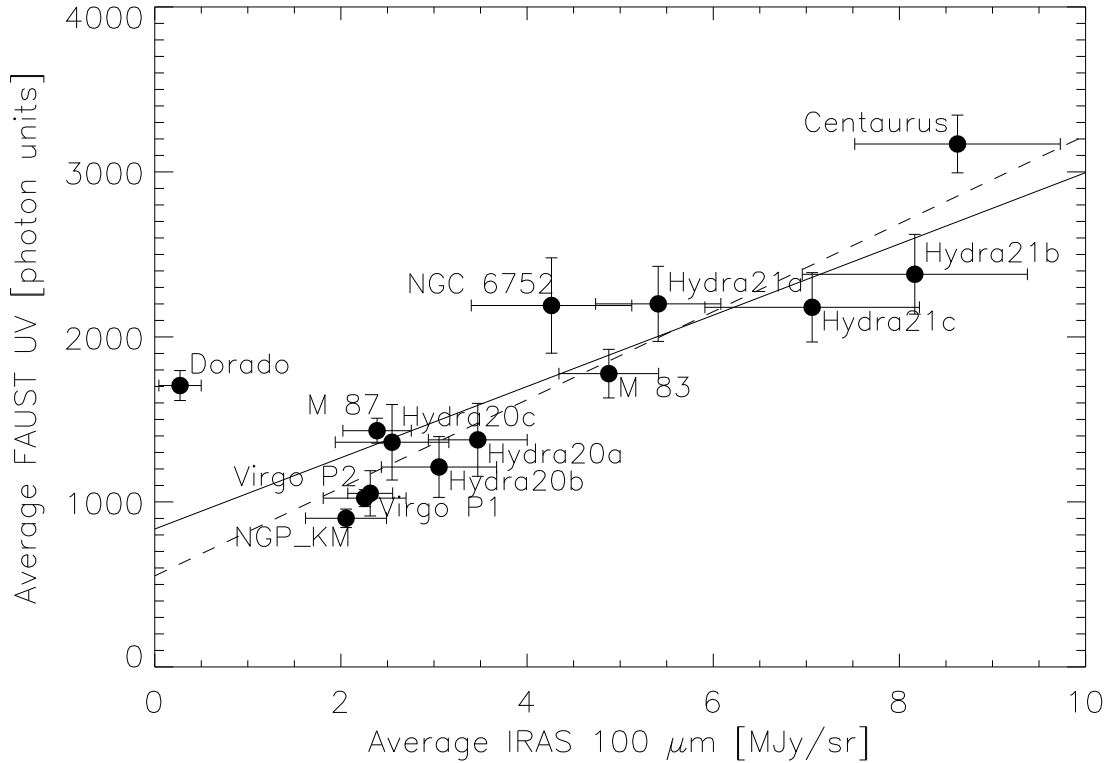}
\caption[f2.eps]{The same UV intensities as in Fig.\ 1 plotted against the
diffuse Galactic IRAS 100$\mu$m background intensity. The solid line indicates the linear
correlation involving all data points; the dashed line shows the linear correlation for the
data with the Dorado measurement excluded. \label{fig2}}
\end{figure}
The IRAS background is correlated both with the dust column density and the density of the
interstellar radiation field responsible for heating the grains along the line of sight.  The
steepening of the correlation suspected in Figure 1 should therefore be reduced; a residual effect
could still result from the decrease of radiation density with z-distance from the galactic plane.  

In Figure 2, the intercept at zero IRAS intensity suggests a background of about 835 units, as 
suggested above.  However, this is rendered somewhat uncertain by three factors: (1) the zero-point
uncertainties (\cite{BP88}) of the IRAS 100$\mu$m detector; (2) the likely presence of an as yet
undetermined extragalactic background component in the 100$\mu$m intensity; and (3) the exceptional 
weight of the Dorado measurement on the slope of the correlation.  Since the high UV intensity at
Dorado is consistent with expectations from our model (see Figure 5), we suspect that the 100$\mu$m
intensity of Dorado is in error.  Eliminating Dorado from the correlation in Figure 2 (dashed line),
we arrive at an intercept of 550 units, in essential agreement with the intercept in Figure 1.  The
closeness of the correlation between DGL and far-IR background in Fig. 2 is marginally better (r~=~0.93,
Dorado omitted) than the correlation between DGL and HI in Figure 1 (r~=~0.79); this may be understood
through the fact that both DGL and IRAS 100$\mu$m intensities depend directly on the dust column
density, while the DGL-HI correlation involves a possibly variable gas-to-dust ratio.
We must attribute the real width in
the DGL distribution for a given IRAS intensity as resulting from variations due to phase
function effects in the far-ultraviolet, 
which would leave the IRAS background unaffected, and from deviations of the
line-of-sight cloud spectrum from the average (see \S 3.3.4).

     This initial examination of the FAUST data revealed that the measured diffuse
background intensities consist of a constant component of at least 600 units and a DGL
component which is correlated with other measures of the galactic dust column density and with
the galactic radiation density.  Provided the first component is, on average, the same for any
random subset of the data, the latter component can be analyzed with an appropriate radiative
transfer model.

\section{The Radiative Transfer Model}
\subsection{General Requirements}

     A successful model for the DGL must have the following properties:
(1) It must contain the actual observed illuminating radiation field,
(2) the scattering medium should have cloud structure, and
(3) multiple scattering must be included.
These requirements become clear when one considers the following points.

     One of the most important characteristics of the DGL at far-UV wavelengths is apparent
from Figure 1:  The DGL intensity for a given N(HI) varies by factors of two-to-three with
direction in the sky after the approximately constant background has been subtracted.  
While part of this variation may be due to a spatially variable 
gas-to-dust
ratio or to dust associated with HII or H$_{2}$ rather than HI, most of the observed variation is a
result of the spatially, severely anisotropic, interstellar radiation field (\cite{Gond90}; 
\cite{HLM88}).  A first requirement of any suitable DGL model for the UV is therefore that it 
must
employ the known distribution of the far-UV sources of radiation.  A radiation field based on
the expected UV fluxes from known hot stars listed in the SKYMAP star catalog (\cite{Got78})
has recently been presented by Murthy \& Henry (1995).  Another useful source is the measured
UV fluxes of some 58,000 stars detected by the TD-1 satellite (\cite{Gond90}), which we have used
for this work.

     The interstellar medium responsible for the opacity of interstellar space is distributed in
the form of structures commonly referred to as clouds.  Clouds containing grains with 
single-scattering albedo $a < 1$ and exposed to an external radiation field will have an albedo less than
the single-scattering albedo (\cite{vdH87}), and this difference in albedo will become 
larger
with increasing optical depth of a cloud and with increasing phase function asymmetry (\cite{WG96}).  
In the far-UV the optical depth per cloud is typically three times that at 
visible
wavelengths, and previous studies suggest an increase of the phase function asymmetry toward
shorter wavelengths (\cite{CBG95}).  Thus, a DGL model suitable for the analysis of far-UV 
data should incorporate a cloud structure for the scattering medium if it is to avoid a
systematic underestimate of the far-UV albedo.

     Van de Hulst and de Jong (\cite{vdHdJ69}) demonstrated that multiple scattering is an indispensable
feature involved in the transfer of DGL.  This applies even more to the UV because of the
greater optical depths of the galactic dust layer.  In the case of the present observations, 
which
are restricted to intermediate and high galactic latitudes with N(HI) $\leq$ 8 x 10$^{20}$ cm$^{-2}$ 
(see Table
1), multiple scattering is marginally important.  With corresponding UV optical depths of order
1.2 or less, the resultant albedo bias can be as high as 25\% if the entire line-of-sight 
material
is contained in a single cloud and a single grain albedo of 0.5 is assumed.  We will, therefore,
include multiple scattering in our model.

\subsection{Model Design}

     The principle of the WP radiative transfer model follows the pioneering approach of
Mattila (1971) by taking into account the discrete cloud structure of the interstellar medium. 
Mattila's treatment was limited to directions in the galactic plane which placed his 
calculations
into the limit of very large optical depths.  This meant that the DGL intensity was determined
mainly by the adopted incident radiation field which was derived from photometric
measurements of the integrated sky brightness as a function of galactic latitude and longitude.

     In the WP model, the approach was generalized to include all galactic latitudes.  As a
source for the (mostly small) optical depths at higher latitudes we adopted the Bell 
Laboratories
HI Survey (\cite{SGW92}), supplemented by the Parkes HI survey (\cite{CHH79}; \cite{HC79}).  
We used the conversion from HI column density to optical depth at the
FAUST effective wavelength of 165nm as given by Sasseen \& Deharveng (\cite{SD96}),
\begin{equation}
                  \tau_{165} = 0.144 N_{H} [10^{20}cm^{-2}],               
\end{equation}
which is based on the N(H) - E(B-V) relationship observed by Bohlin et al.\ (1978)
and Diplas \& Savage (\cite{DS94}), and the average galactic extinction curve of Savage \& Mathis (1979).

\subsubsection{Radiation Fields}

     In order to extend the radiative transfer model to directions with $|b| > 0\arcdeg$,
we need to express the variation of
the radiation field with z-distance from the plane.  We start by constructing
the radiation field in the galactic plane (z~=~0) utilizing the fluxes of about 58,000 
stars
measured by TD-1 at $\lambda$~=~156.5nm (FWHM~=~330\AA) which were kindly provided to us by J.
Murthy.  Maps of this radiation field may be found in the recent work of Murthy \& Henry
(1995) and in the Atlas of the Ultraviolet Sky by Henry et al. (1988).  

     The TD-1 catalog is thought to be complete down to fluxes of 1.0~x~10$^{-12}$ erg cm$^{-2}$
 s$^{-1}$ \AA$^{-1}$ at 156nm.  The contribution by stars fainter than this limit to the 
integrated stellar fluxes appears to be negligible at this wavelength, certainly much less than 
10\% of the total (Gondhalekar 1990).  However, the DGL observed by TD-1 (\cite{MNT78}) is part
of the illuminating radiation field seen by each cloud and needs to be added to the integrated
stellar flux.  We therefore computed the expected DGL radiation field for a dust albedo $a = 0.5$
and a strongly forward directed phase function and added it to the integrated stellar fluxes.
This increased the total background intensities by typically 25\%.

  For the purpose of our
calculation, we represented this radiation field (at z = 0) as an 800-element, two-dimensional
array, and expressed the average intensity in each array element in photon units.  The spacing
of the array elements and their sizes are non-uniform.  At low galactic latitudes, and other
directions where large gradients in the radiation field intensity occur, elements are spaced 
more
closely, while at high galactic latitudes large elements are chosen.

     The transformation of the central-plane radiation field into one seen by a cloud located
at distance z$_{C}~\neq$~0 from the galactic plane was approximated following a method described by
Mattila (\cite{Matt80a},\cite{Matt80b}). For the purpose of this transformation only, we assumed 
that the principal source of the $\lambda$~=~156.5nm radiation field is
galactic OB stars whose volume emissivity is distributed in a plane-parallel disk with an
exponential z-distribution characterized by a vertical scale height $\beta$~=~60 pc.  
We assume this because
about 90\% of the total integrated stellar radiation at 156nm arises near the galactic plane defined
by $-45 \arcdeg \leq b \leq 45 \arcdeg$ (Gondhalekar 1990).
Also, this assumption appears justified because the number of suspected subdwarfs in the TD-1 catalog,
though greater than anticipated, was only of order 10$^{3}$ (\cite{CW83}).  The galactic distribution
of these stars does not differ drastically from that of the overall stellar radiation field and their
contribution to the radiation field was included in any case. 
By comparing the
expected $b$-distribution of the integrated intensity seen by an observer at z~=~z$_{C}$ with the
symmetric distribution seen at z = 0, we derived transformation factors as a function of z$_{C}$ and
$b$.  This was done by dividing the skewed radiation field distribution at z~=~z$_{C}$ by the
symmetric distribution at z~=~0.  To obtain an actual 
radiation field applicable to z~=~z$_{C}$, we then multiplied the TD-1
radiation field with the appropriate $b$-dependent transformation factors.  As a result, radiation
fields at increasing $|$z$_{C}|$ exhibit a growing asymmetry in the radiation emanating from the two
galactic hemispheres with $b>0\arcdeg$ and $b<0\arcdeg$, and the near-equatorial intensity peak shifts toward
latitudes in the opposite hemisphere while maintaining the basic non-isotropy and longitudinal
variability of the original TD-1 field.  We constructed a total of 17 radiation fields, for 
values
of z$_{C}$~=~0, $\pm$25pc, $\pm$50pc, $\pm$75pc, $\pm$100pc, $\pm$150pc, $\pm$200pc, $\pm$250pc, 
and $\pm$300pc.

\subsubsection{Cloud Spectrum}

     Interstellar clouds exist with a wide range of optical depths (e.g. Crovisier 1981).  In
order to reflect these conditions we adopted a cloud spectrum of three types of spherical 
clouds: 
uniform clouds of optical depth diameter (at 156.5nm) of 0.5, uniform clouds of optical depth
diameter of 2.5, and centrally condensed clouds ($\rho \propto$ r$^{-2}$ for r/R $\geq$ 0.05;
$\rho$ = constant for r/R
$<$ 0.05) of optical depth diameter of 10.  The probability ratio of encountering such clouds
along any line of sight was taken to be 12:3:1.  Assuming arbitrary impact parameters for a
penetrating line-of-sight, such a cloud spectrum provides, on average, a column density
distribution very similar to that derived by Crovisier (\cite{Cro81}) from the statistical properties of
interstellar HI which were based on 21-cm line absorption surveys.  The average optical depth per cloud for our
adopted spectrum was found to be 0.593 by subjecting our spectrum of three clouds, in a frequency ratio of
12:3:1, to random penetration by a line of sight.  The Crovisier study seemed to be particularly relevant
in the current context:  (1) its database consists mainly of observations out of the galactic 
plane,
as is the case with the FAUST data; (2) our source for the column density of dust is the
correlation between E(B-V) and HI (\cite{BSD78}); (3) the range of total HI
column densities covered by the Crovisier study coincides with the range encountered in the
FAUST fields in this study.

     The probability of encountering a cloud of any type at a distance z from the galactic
plane was assumed to be given by an exponential distribution with a scale height of 110 pc, i.e.
was assumed to be proportional to exp (-z/110 pc) (Crovisier 1981; \cite{Lock84}).

\subsection{Model Operation}

     The computation of the expected DGL surface brightness in a given direction with our
model involves several Monte Carlo processes:  the multiple scattering radiative transfer within
each cloud of our cloud spectrum, the determination of the number of clouds in a given
direction, the determination of the type and location of the cloud along the line of sight, the
determination of the individual surface brightness profile of each cloud along the line of 
sight,
the determination of the penetration point of each cloud, the determination of the optical depth
along the chosen line of sight through each cloud, and finally the integration of intensities of 
the
clouds along the line of sight.

\subsubsection{The Radiative Transfer in Individual Clouds}

     The transfer solution for a spherical cloud of given optical depth diameter, dust 
scattering
parameters given by albedo $a$ and phase function asymmetry $g$, and density profile was found
with the method outlined in Witt \& Stephens (1974) and in FitzGerald et al. (\cite{FSW76}).  The
method traces the transfer of individual photons in reverse and produces weight matrices which
express the probability that photons hitting the cloud from any direction will exit the cloud at
a given projected radius in the direction of the observer.  For the non-isotropic scattering 
phase
function, the Henyey-Greenstein (\cite{HG41}) function was chosen.  Our approach has the advantage
that the radiative transfer through individual clouds needs to be done only once for a given set
of parameters.  The specific surface brightness profile exhibited by a cloud at a given z$_{C}$ can 
be
obtained by multiplying the cloud's probability matrix with the appropriate radiation field 
matrix
corresponding to the value of z$_{C}$.

\subsubsection{The Number, Type, and Location of Clouds along the Line of Sight}

     The average number of clouds along a given line of sight with total column density N(HI) is
found by dividing the corresponding optical depth derived from Eq. (1) by the average optical
depth per cloud $<\tau_{C}>$~=~0.593 (\S 3.2.2).  For a typical line of sight in the galactic plane
(A$_{V}$ = 1.9 mag/kpc; \cite{Spit78}), with an average ratio of N(HI)/E(B-V)~=~4.8~x~10$^{21}$ 
atom cm$^{-2}$ mag$^{-1}$ (\cite{BSD78}), we find 7.1 clouds/kpc, 
in close agreement with estimates discussed by Spitzer (1978).  For typical high-$|b|$ lines of sight,
we generally find 2, 1, or no clouds.

    The actual number of clouds, n, for a given Monte Carlo
simulation is then selected by a Monte Carlo process from a Poisson distribution whose average
is the average number of clouds just determined.  The types of clouds (optical thickness
diameter, density distribution) are found by random-number choices from within our cloud
spectrum based upon their relative frequencies.  The z-distances of these clouds are then 
derived
from yet another series of Monte Carlo processes applied to the assumed exponential distribution
of cloud distances from the Galactic plane.  This sampling of cloud distribution by number,
types, and location along the line of sight was performed 300 times for each of the
approximately 60 pixel positions within each FAUST field for a total
of 1.8~x~10$^{4}$ integrations per field.  Thus, the statistical properties of
the interstellar medium are interwoven into the prediction of the DGL intensity.  We therefore
predict an actual distribution of expected DGL intensities for each pixel, whose FWHM provides
a measure of the DGL uncertainty resulting from possible deviations from average conditions.
Examples of such distributions are shown in Figure 3 and will be discussed below in \S 3.3.4.

\subsubsection{Surface Brightness Profiles and Total DGL Intensities}

     The probability weight matrix of each of the randomly selected and positioned clouds
(\S 3.3.2) must be multiplied by the appropriate radiation field in order to predict their 
radial
surface brightness profiles.  The most abundant variety of clouds, the optically thin ones, 
exhibit
a brightness profile with a maximum at the projected center; the moderately optically thick
clouds present a fairly flat surface brightness profile with limb darkening, while the highly
optically thick clouds display a dark central region surrounded by a bright rim (\cite{WS74}).  
The surface brightness value added by each cloud along the line of sight is determined
by the selection of a random penetration point on the geometric cross section of a cloud.  In
order to correct the surface brightness of still more distant clouds seen in this direction, we 
also
determine the optical depth through the entire cloud at the penetration point.  If the 
intensities
of the clouds along a given direction are denoted by I$_{1}$, I$_{2}$, I$_{3}, \ldots,$ 
the total integrated intensity 
for
this line of sight (\cite{Matt71}) is
\begin{equation}
      I_{DGL} = I_{1} + I_{2}e^{-\tau_{1}} + I_{3}e^{-(\tau_{1}+\tau_{2})} + \ldots    
\end{equation}

\subsubsection{Dependence of Results on Model Parameters}

     In its attempt to represent the DGL radiative transfer in as realistic a manner as 
possible,
the model described above involves numerous parameters.  While most of these, such as the
radiation field distribution and the optical depth distribution, are well constrained by 
available
observations and may therefore be considered fixed, it is the nature of the adopted cloud
spectrum that affects the model predictions most profoundly.  As indicated in \S 3.1, the
greater the optical thickness of an individual cloud, the lower its resultant effective albedo 
for
a given radiation field and single grain albedo.  In addition, in directions with low column
densities, a cloud spectrum incorporating clouds of higher optical depth naturally leads to
instances of "holes", i.e. directions with no clouds and thus, zero DGL intensity, because the
beam-averaged optical depth must still be that constrained by the HI column density.  If, on the
other hand, the cloud spectrum consists entirely of very-low-$\tau$ clouds, there will always by
numerous clouds along lines of sight with small overall column density, assuring a more uniform
sky coverage with more efficiently scattering clouds.

     We illustrate these effects in Figure 3 by showing the results of 100 samplings with 300
photons each along a single direction ($\ell = 315\arcdeg, b = +35\arcdeg$, N(HI) = 5.15 x 10$^{20}$cm$^{-2}$)
for three different cloud spectra.  
\begin{figure}[htb]
\plotone{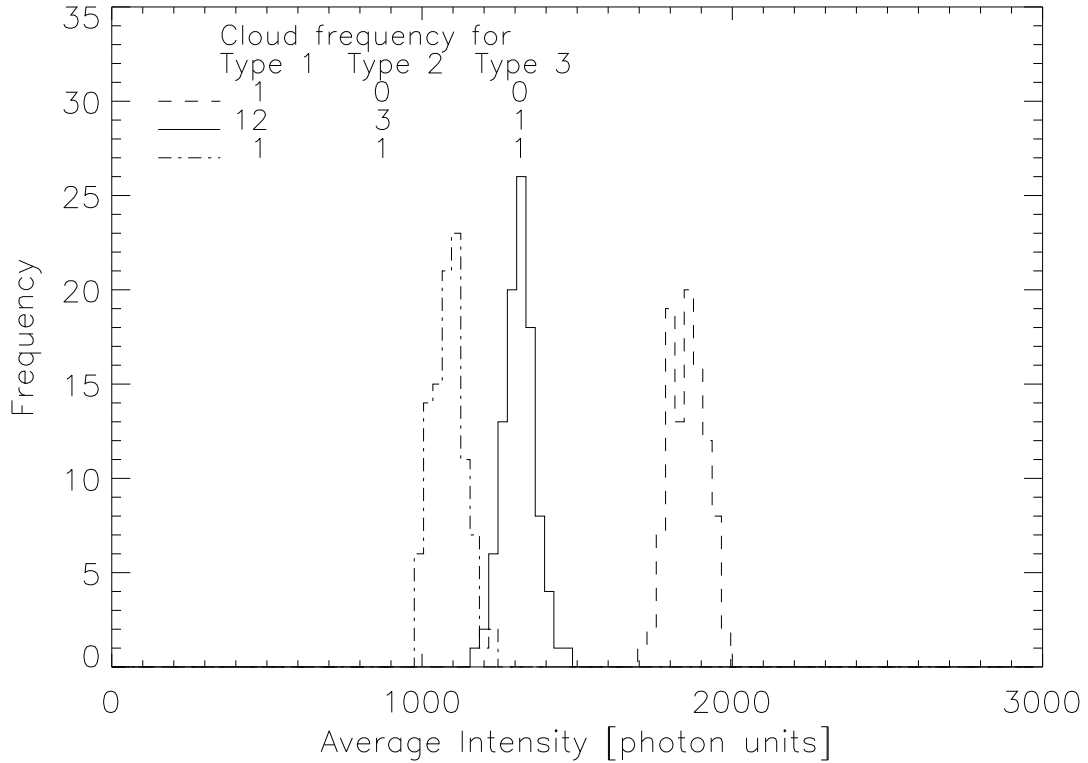}
\caption[f3.eps]{Model prediction of the intensity of UV DGL plus UV background
derived from 100 samplings with 300 photons each for three different cloud spectra.  All
calculations are for the identical direction ($\ell~=~315\arcdeg$, $b~=~+35\arcdeg$) and
for identical grain properties ($a~=~0.45$, $g~=~0.675$).  The hydrogen column
density is N(HI)~=~5.15~x~10$^{20}$cm$^{-2}$.    \label{fig3}}
\end{figure}
The first consists of optically thin clouds of type 1 only, with an optical
depth diameter of 0.5 at 156.5nm wavelength.  The average optical depth for cloud penetration
is 0.33.  The predicted DGL intensity is 1871~$\pm$~59 units.  The second is our adopted case of
a cloud spectrum with ratios 12:3:1 divided among clouds of type 1, 2, and 3, respectively.  The
average optical depth per cloud is now 0.59 and the predicted DGL intensity is 1329~$\pm$~52 units,
some 30\% lower than case 1.  The third case displayed is for a cloud spectrum 1:1:1, which
now emphasizes optically thick clouds.  The average optical depth per cloud is 0.83 and the
predicted DGL, as expected, is still lower at 1101~$\pm$~52 units.  All calculations assumed the
same scattering characteristics, namely $a = 0.45$ and $g = 0.675$.

     For the analysis of the FAUST data we have adopted a cloud spectrum which is
dominated by low-$\tau$ diffuse clouds, but also contains a small number of clouds of greater optical
thickness in a ratio of 12:3:1 in terms of their average relative frequency along a given line
of sight.  This spectrum reflects the demonstrated presence of numerous small molecular clouds
at high galactic latitudes (\cite{MBL86}; \cite{MHS96}; \cite{Stark95}; see \cite{Mag94}
for a recent review).  Our adopted spectrum is also consistent with the observed
distribution of HI column densities based on the 21-cm line absorption survey by Crovisier (\cite{Cro81}). 
It should be clear from Fig. 3 that the albedo derived from the FAUST data using our adopted
cloud spectrum will be higher by about 30\% than it would be, had we assumed low-$\tau$ clouds
entirely.  The dependence of the intensity of the predicted DGL upon the chosen cloud spectrum
must also be kept in mind when we compare predicted DGL intensities with those actually
observed (see Fig. 6).

\section{Derivation of Scattering Properties}

     Our aim in this section is two-fold:  (1) We want to separate the galactic component of
the diffuse UV intensities measured by FAUST (Table 1) from the non-galactic components. 
Contributions to the latter are terrestrial airglow and extragalactic background radiation.  (2) 
We want to analyze the galactic component using the DGL model described in the previous section. 
The aim is to derive the average dust albedo $a$ and the average phase function asymmetry
parameter $g$ which characterize the galactic dust in the diffuse ISM at intermediate and high
galactic latitudes.  Toward this end we calculated the expected average UV intensity for each
of the 14 FAUST fields, each based on approximately 1.8~x~10$^{4}$ integrations, for a wide grid of
closely spaced values of $a$ and $g$.

\subsection{The Fitting Criteria}

     Which set of scattering parameters for dust grains will lead to the best representation of
our data?  To decide this question, we assumed the following.  (a) All FAUST lines of sight
contain dust with identical average properties.  Since for our observations $|b|>21\arcdeg$, most of the
clouds studied are well outside the galactic plane and are representative of the low-density diffuse
interstellar medium.  (b) The measured surface brightness of each FAUST field contains an
unknown component with a surface brightness $\geq$~0 units due to residual airglow and a
component of extragalactic origin.  We assume the latter to have an intensity of 300 units 
(WP \cite{WP94}) outside the Galaxy.  This component is assumed to be isotropic but to be
attenuated for each line of sight by the galactic dust column derived from the N(HI) values.  We
included this extragalactic component both into the illuminating radiation field (essentially
negligible) and into the predicted intensity for each FAUST field.  (c) On average, the FAUST
fields at lower galactic latitudes are subject to the same airglow contamination as the FAUST
fields at higher galactic latitudes.  This does not require that each field have an identical 
airglow
component, only that any random group of six or seven fields have similar airglow averages. 
Thus, the galactic latitude dependence seen in our measured intensities is due entirely to DGL
and the extinction-modulated extragalactic background with the former generally increasing with
decreasing latitude and the latter decreasing with the increasing extinction at lower latitudes.
     Given these assumptions, the best-fit model must satisfy these requirements:
\begin{enumerate}
\item Slope Criterion:  The model predictions of the DGL values for the FAUST fields,
          including the partially attenuated extragalactic radiation, must exhibit the same
          slopes as the observations when plotted against the corresponding values of
          N(HI), cosec$|b|$, and the IRAS 100$\mu$m intensities for the FAUST fields.
\item Positive Airglow Criterion:  The average predicted intensities of DGL and
          extragalactic light for individual FAUST fields should in no case fall
          above the measured intensities since this would imply negative airglow.
\item Equal Airglow Criterion:  A considerable range of dust scattering parameters
          ($a,g$) will be able to satisfy the first two criteria.  However, for most sets of
          values of ($a,g$), the value of the positive airglow derived from application of
          criteria (1) and (2) is different when observations and model predictions are
          plotted against N(HI) and when plotted against cosec$|b|$.  The equal airglow
          criterion demands that the value of the airglow deduced from the comparison of
          intercepts is the same when observations and model predictions are plotted
          against N(HI) and cosec$|b|$ and evaluated at N(HI)~=~0 and cosec$|b|$~=~0. 
\end{enumerate}
    
	A much more limited set of values ($a,g$) will satisfy this third criterion in addition to
satisfying the first two.  We do not include the intercept in the plot against the IRAS 100$\mu$m
diffuse intensity in this criterion because calibration uncertainties in the zero point of the 
IRAS
intensity scale and the presence of some extragalactic component in the IRAS data would render
the outcome uncertain.

\subsection{Models}

     In Figure 4, we show all combinations of $a$ and $g$ for which complete models were
calculated.  
\begin{figure}[htb]
\plotone{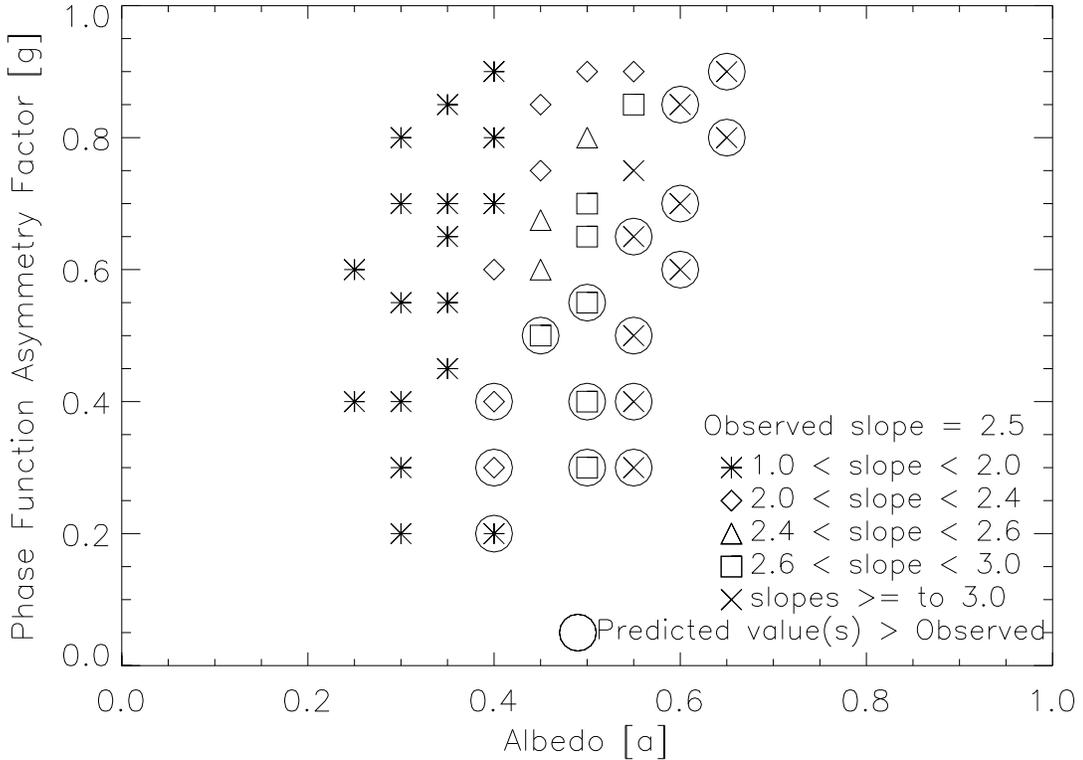}
\caption[f4.eps]{Complete models were calculated for the grain scattering properties
indicated by the symbol locations in the ($a,g$) plane.  Different symbols indicate the slopes
in graphs of the predicted UV intensities vs.\ N(HI).  The encircled symbols correspond to
models which predict intensities higher than observed.  \label{fig4}}
\end{figure}
Different symbols indicate the slope found when predicted intensities are plotted
against the N(HI) values for the corresponding FAUST fields.  The observed slope (see Fig. 1)
is 2.5~[photon units/10$^{18}$cm$^{-2}$].  It is clear from Fig. 4 
that the fitting criterion (1) is met within a very
narrow range of $a$-values (0.4~$<~a~<$~0.5) but for a large range of $g$-values.  The positive
airglow criterion eliminates all models identified by encircled symbols, as well as all those 
with
still larger $a$-values which were left uncomputed.

     As an example, a model with $a$~=~0.45 and $g$~=~0.675 is shown in Fig. 5 in comparison
with the FAUST data plotted against the N(HI) values of the corresponding FAUST fields.  
\begin{figure}[htb]
\plotone{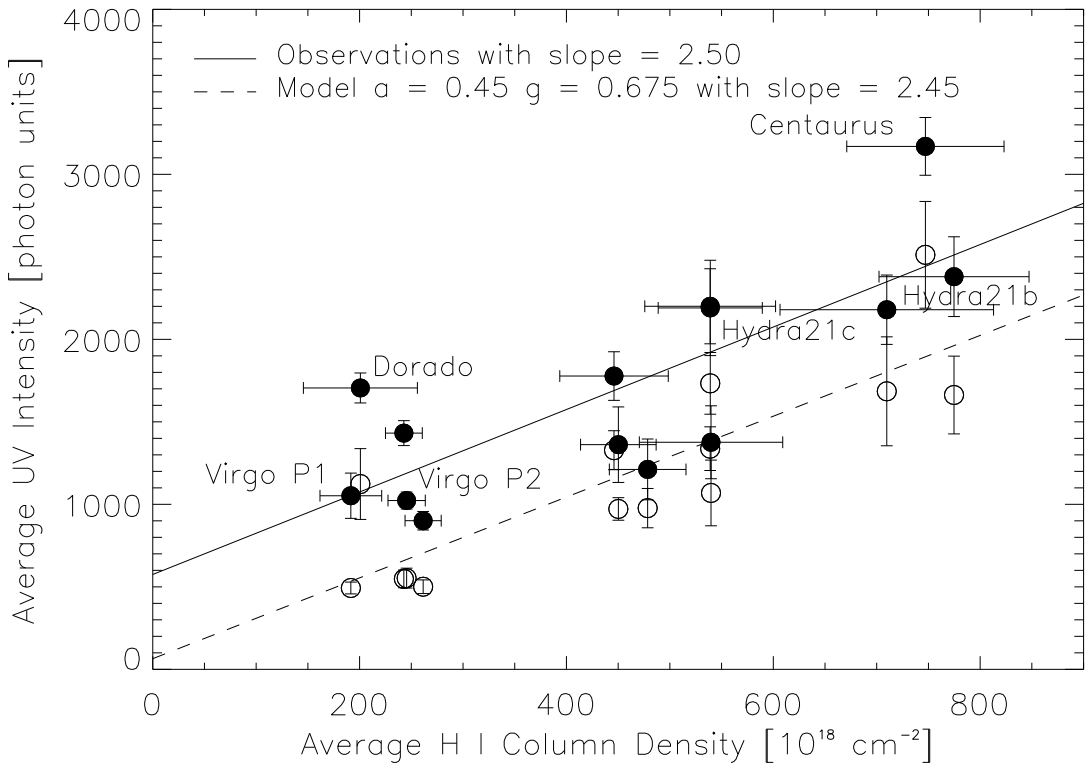}
\caption[f5.eps]{Model predictions (open circles) are plotted together with 
corresponding observations against average N(HI) values for FAUST fields.  The vertical model
error bars denote one standard deviation in the distribution of predicted intensities
for a 4$\arcdeg$ x 4$\arcdeg$ field.  \label{fig5}}
\end{figure}
The model points, denoted by open circles with vertical error bars, follow the observed intensities
closely, except they are displaced by a constant vertical offset attributed to the airglow
component.  For example, the model successfully reproduces the significant differences in
intensity observed for groups of FAUST fields for which the HI column density is almost
constant.  Specific cases are the groups consisting of Hydra 21c, Centaurus and Hydra 21b, and
Virgo P1, Dorado, and Virgo P2.  This gives us confidence both in the model's predictive
power and in our assumption that the residual airglow contribution is roughly constant.  Note
that the positive airglow criterion (2) is also met in the model shown in Fig. 5.

     To further illustrate the close match between the model predictions with the individual
observations, we added an average airglow intensity of 530 units to the model values and
overplotted these on the observations in Figure 6. 
\begin{figure}[htb]
\plotone{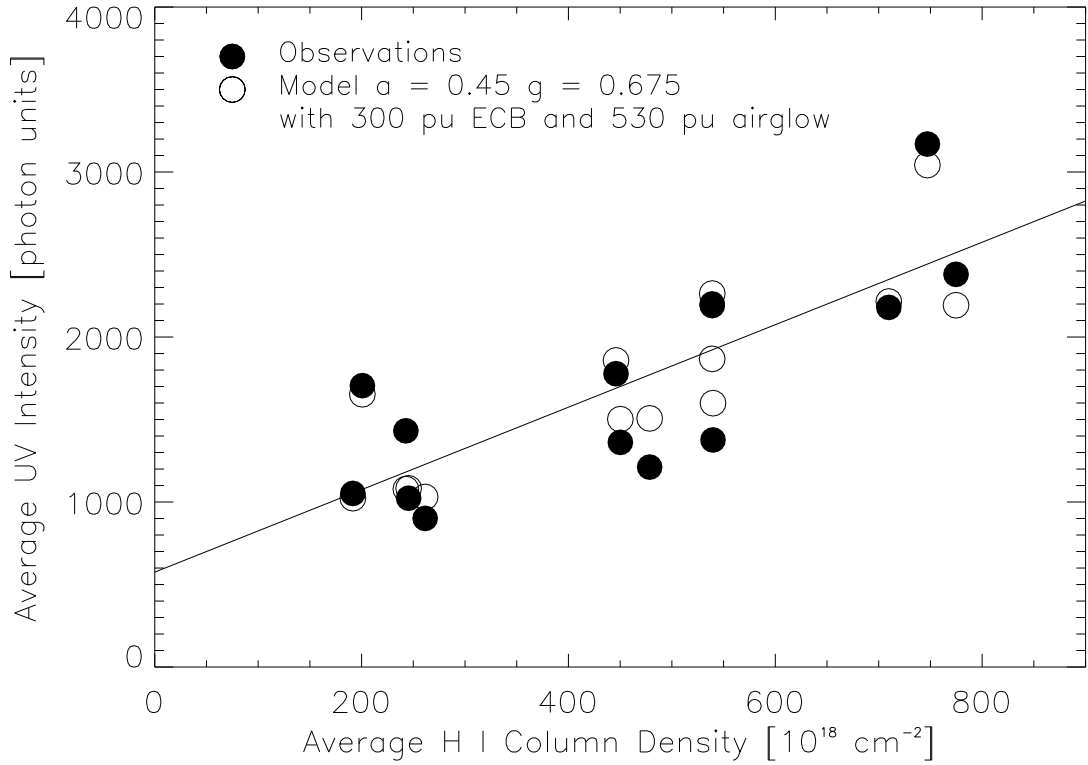}
\caption[f6.eps]{The same data and model predictions as in Figure 5, except that a 
constant airglow intensity of 530 units has been added to the model points.  Error bars have
been omitted for clarity;  they would be identical to those shown in Figure 5.
Note that the observed points for NGC 6752 and Hydra 21a essentially coincide.  The
corresponding model points are those immediately above and below the observed points.   \label{fig6}}
\end{figure}
In almost all cases, the observations are
reproduced within better than one standard deviation.

\subsection{The Equal Airglow Criterion Applied}

     Thus far we have found combinations of $a$ and $g$ which reproduce the slope of the observed
data in Figure 5.

     As mentioned, we interpret the vertical offset between the least-squares slopes fitted to
the observed points and the predicted points in Fig. 5 as resulting from airglow.  The airglow
intensity derived from Fig. 5 is about 530 units, on average.  When the same data and
predictions are plotted against cosec$|b|$ of the positions of the FAUST fields, we arrive at Fig.
7.  
\begin{figure}[htb]
\plotone{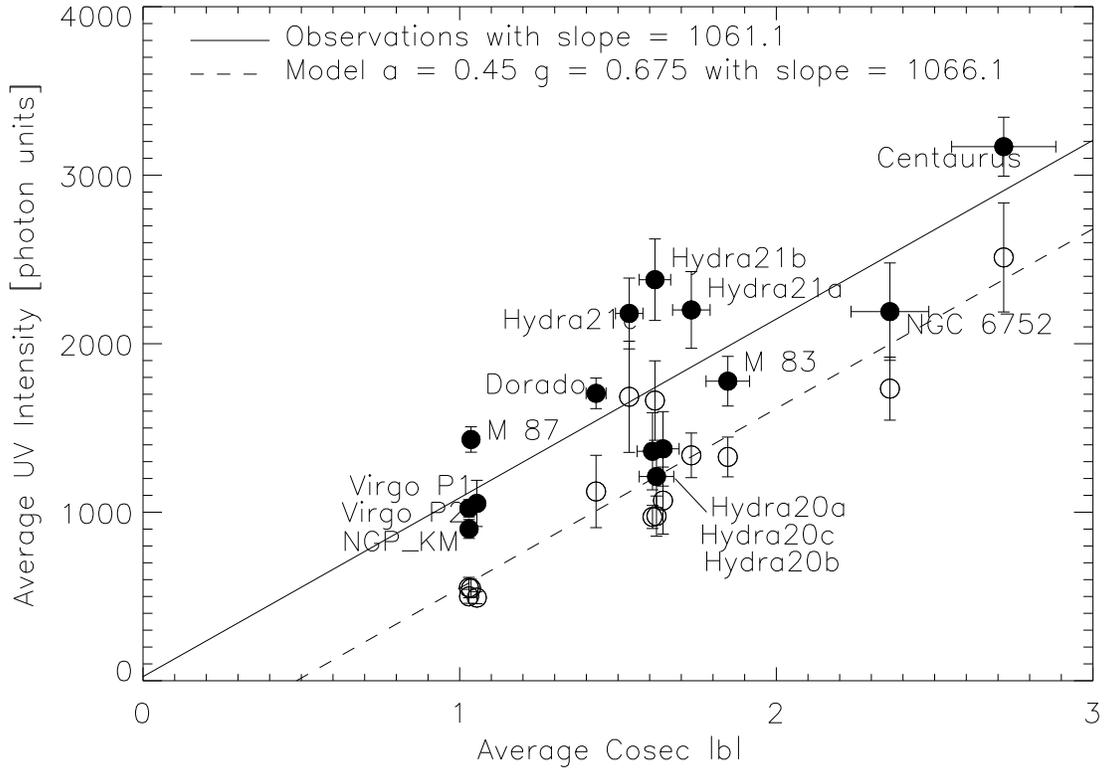}
\caption[f7.eps]{The same data and models as in Figure 5 plotted against cosec $|b|$
of the FAUST fields.  The ``best fit'' solution shown here requires that the vertical offset
between observed slope and model slope be the same as in Fig.\ 5.   \label{fig7}}
\end{figure}
Again, the observed and the model slopes are nearly identical and the vertical offset between
the two slopes is approximately 530 units, i.e., the inferred airglow intensity is the same as
derived from Fig. 5.  If the $g$-values of the models are increased beyond 0.675, the cosec$|b|$
plot yields higher airglow values than the N(HI) plot as well as divergent slope values; if $g$ is
reduced below 0.675, the reverse occurs.  This is shown quantitatively in Fig. 8 where lines
of constant airglow as derived from the two types of plots are shown in the relevant area of the
($a,g$) plane.  Allowable values for $a$ and $g$ must be located in the part of the plane where
the two comparisons yield the same airglow values.  
\begin{figure}[htb]
\plotone{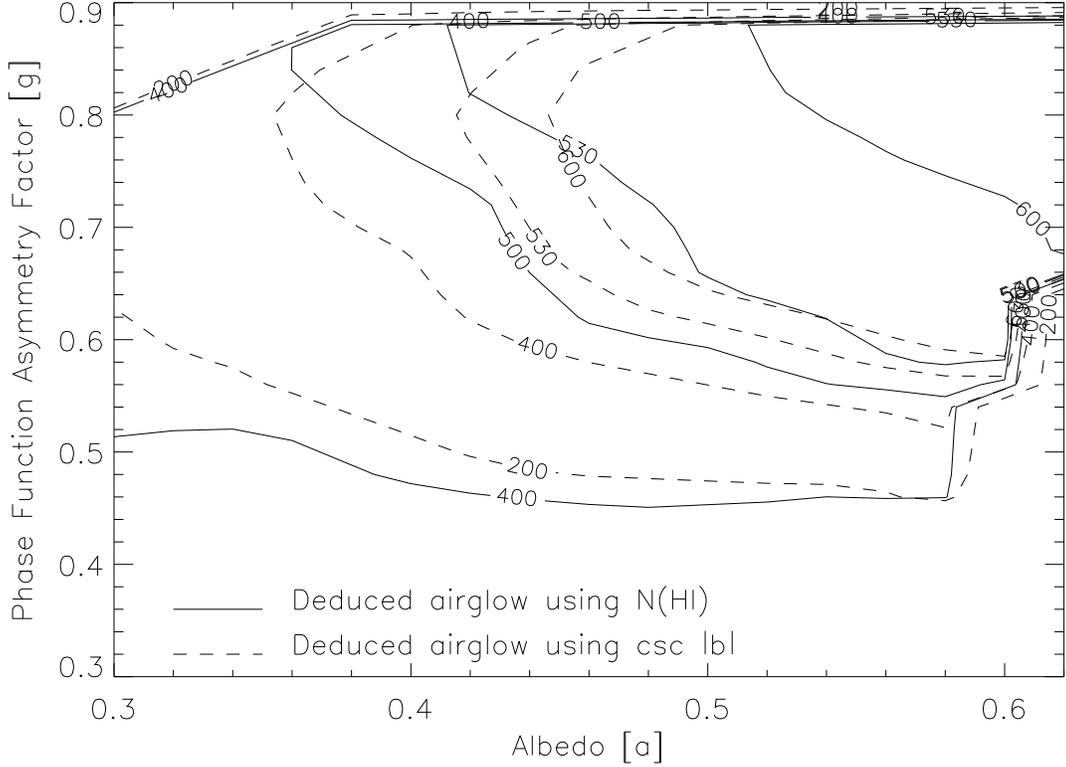}
\caption[f8.eps]{Contours of constant airglow deduced from comparing models and 
observations when plotted against N(HI) and against cosec$|b|$.  This graph suggests an average
airglow near 510 photon units as being associated with the ``best fit'' solution.   \label{fig8}}
\end{figure}

     The straight forward application of the three fitting criteria constrains the best solution
to a narrow region surrounding the point $a$~=~0.45 and $g$~=~0.675 in the ($a,g$) plane.  Using
this solution, we have derived the likely airglow intensities for our 14 FAUST fields and have
listed them in Table 2. 
\begin{deluxetable}{lc}
\tablewidth{210pt}
\tablecaption{Deduced Airglow Values\tablenotemark{1} \label{table2}}
\tablehead{
\colhead{FAUST Field} & \colhead{Deduced Airglow} \\
\colhead{} & \colhead{(Photon units)}
}
\startdata
Dorado & 582 $\pm$ 233 \nl
NGPKM & 400 $\pm$ \phn 70 \nl
M 87 & 883 $\pm$ \phn 93 \nl
Virgo P2 & 470 $\pm$ \phn 79 \nl
Virgo P1 & 559 $\pm$ 142 \nl
Centaurus & 657 $\pm$ 368 \nl
M 83 & 449 $\pm$ 368 \nl
NGC 6752 & 457 $\pm$ 344 \nl
Hydra 20a & 307 $\pm$ 298 \nl
Hydra 20b & 307 $\pm$ 220 \nl
Hydra 20c & 390 $\pm$ 239 \nl
Hydra 21a & 863 $\pm$ 263 \nl
Hydra 21b & 717 $\pm$ 338 \nl
Hydra 21c & 494 $\pm$ 391 \nl
\enddata
\tablenotetext{1}{Based on the assumption that the extragalactic isotropic 
background before galactic extinction is 300 photon units.}
\end{deluxetable}

\subsection{Uncertainties}

     It was shown in Fig. 4 that the albedo value of the "best fit" solution is sensitively
dependent upon the slope of the measured FAUST diffuse intensity plotted against N(HI). 
Because of the close correlation between N(HI) and cosec$|b|$ and between N(HI) and the IRAS
100$\mu$m diffuse galactic background (\cite{BP88}; \cite{BAB96}), the same
would be the case, if plots are made against either of these two other quantities.

    There are two major sources of uncertainty in these observed slopes:  (1) The residual
airglow is variable with time and direction, and thus, a randomly variable component is added to
the true DGL values.  The values in Table 2 are representative of the range of variability to be
expected for the airglow component.  Only if our DGL model were perfectly accurate would
Table 2 give the actual airglow values, and we do not assume this.  The presence of a random
airglow component having the same distribution as Table 2 among our data renders the slope of
the linear regression in the plot against N(HI) uncertain
by $\pm$~0.34 [photon units/10$^{18}$cm$^{-2}$].  (2) The
absolute calibration of the FAUST instrument is uncertain by about 15\% (\cite{BSLW93}). 
While this does not affect the uncertainty of the linear regression, it affects each data point 
by the same percentage, and thus both the intercept and the slope are uncertain.  We find the 
impact of the systematic calibration uncertainty 
upon the slope to be $\pm$~0.37 [photon units/10$^{18}$cm$^{-2}$].   Adding
these uncertainties in quadrature, we find for the 
observed slope 2.5~$\pm$~0.5 [photon units/10$^{18}$cm$^{-2}$]. 

     An additional uncertainly arises from the fact that two independent model calculations
with identical parameters but different seeds for the random number generator produce model
slopes which typically vary by $\pm$~0.10 
[photon units/10$^{18}$cm$^{-2}$], which must be added to the uncertainty
above when observations and models are compared.

\subsection{Scattering Properties}

     We again plot in Fig. 9 the ($a,g$) plane as in Fig. 4 in which the region of possible ($a,g$)
values is now delineated by the application of the three fitting criteria and the uncertainties 
in the slope discussed above.  
\begin{figure}[htb]
\plotone{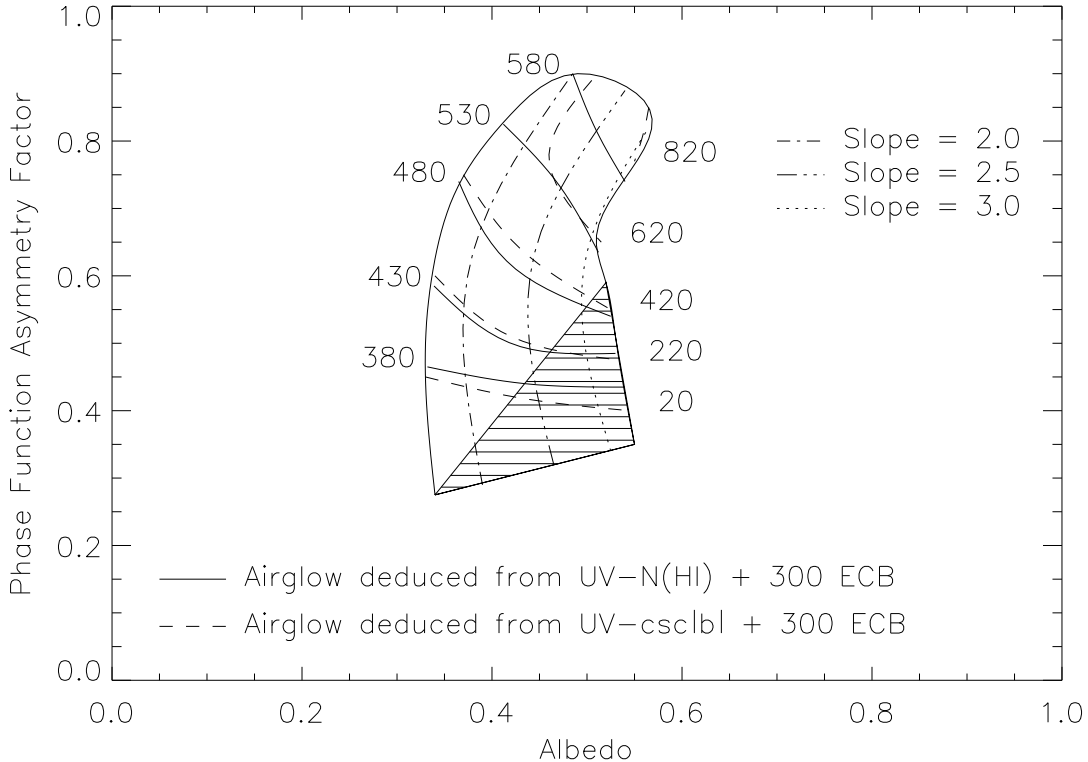}
\caption[f9.eps]{The derived scattering properties in the ($a,g$) plane as constrained
by the three fitting criteria (\S 4.1).  The slopes referred to are those of the predicted
UV-DGL-N(HI) relationships.  \label{fig9}}
\end{figure}
The outer boundary is set by the slope uncertainty in the plot of
FAUST intensities against N(HI).  The cross-hatched area is considered unlikely for possible
solutions since the model, with no airglow, overpredicts the observed intensity.  The equal
airglow criterion and the slope criterion lead to a most probable value for the albedo, $a$~=~0.45
~$\pm$~0.05, and for the phase function asymmetry, $g$~=~0.68~$\pm$~0.10.

\section{Discussion}
\subsection{Scattering Properties}

     Ultimately, we would like to know whether the population of dust grains is different in
different astrophysical environments within the Galaxy.  Standard dust grain models are based
mainly on the observation of the average galactic extinction curve (Savage \& Mathis 1979) and
observed depletions of heavy elements from the gas phase (\cite{SCS94}; \cite{SW95},
\cite{SW96}), and are therefore representing mostly the diffuse interstellar medium near the Galactic
plane.  However, differences in extinction characteristics between different lines of sight 
(\cite{WBS84}; \cite{FM90}; \cite{MC92}; \cite{JG93})
as well as differences in depletion patterns (\cite{SS96}) are well documented, and
these are reflected in differences in the size distributions of grains derived for different
environments (e.g. \cite{KMH94}).

     Scattering characteristics, like extinction properties, are 
also dependent on grain composition and size.  There
is little reason to suspect that dust grains in different environments, e.g. reflection nebulae, 
HII regions, or the low-density diffuse ISM, are similar with respect to their scattering 
properties.  This is 
especially so in light of the considerable observational evidence that the extinction properties of 
the
dust exhibit substantial variations in the UV, both in magnitude and in wavelength dependence.

     In Table 3 we have compiled a number of values of ($a,g$) which have been derived
\begin{deluxetable}{cccccc}
\tablecaption{ \label{table3}}
\tablehead{
\colhead{Region} & \colhead{$\lambda$ (nm)} & \colhead{$a$} & \colhead{$g$} & 
\colhead{R$_{V}$} & \colhead{Ref}
}
\startdata
NGC 7023 & 144,152 & $\geq 0.65 \pm 0.09$ & 0.75 $\pm$ 0.05 & $\sim$ 4 & 1,5 \nl
NGC 7023 & 130 & 0.62 $\pm$ 0.09 & 0.75 $\pm$ 0.05 & $\sim$ 4 & 2,5 \nl
Sco OB2 Assn & 177 & 0.64 $\pm$ 0.09 & undet. & 3.5 & 3 \nl
IC 435 & 156 & 0.75 $\pm$ 0.05 & 0.7 $\pm$ 0.1 & 5.3 & 4,6 \nl
Diffuse ISM & 160 & 0.45 $\pm$ 0.05 & 0.68 $\pm$ 0.1 & 3.1 & this study \nl
\enddata
\tablerefs{1. \cite{WPB92}; 2. \cite{WPH93}; 3. \cite{GWC94}; \\
4. \cite{CBG95}; 5. \cite{WYF80}; 6. \cite{SW89}}
\end{deluxetable}
recently from the observation and analysis of scattered UV light in reflection nebulae and star
forming regions, all regions characterized by densities in the range 10$^{2}~$cm$^{-3} <$ n(H) $\lesssim$
 10$^{4}$~cm$^{-3}$. 
The denser reflection nebulae appear to share albedo values in excess of 0.6 in the 130-170nm
spectral range, while the present study yields a value definitely lower than that.  This 
difference
is most easily understood if we assume that the average grain size is larger in reflection 
nebulae,
or, what would amount to the same, that the number of very small grains (size~$\ll \lambda$) is
reduced compared to the number of larger (size~$\sim \lambda$) grains.  This is indeed supported by the
fact that in reflection nebulae R$_{V}$~=~A$_{V}$/E(B-V) is usually found to be larger than the "normal"
galactic value for the diffuse ISM of R$_{V}$~=~3.1, which is true also for the regions compared in
Table 3.  In addition, most of the dust observed in this study is seen at relatively high 
galactic
latitudes, placing it potentially at large z-distances. With our assumed z-distribution,
50\% of the dust is located at $|$z$|~>~75$ pc. 
Sembach and Savage (1996) have shown that
dust grains in such environments are subject to erosive processes which return a portion of 
their
solid substance to the gas phase.  Kim and Martin (\cite{KM96}) have shown that size distributions of
interstellar grains characterized by R$_{V}~>$~3.1 have higher albedos than the standard R$_{V}$~=~3.1
grains throughout the entire near-IR to UV spectral range.  There are still discrepancies,
however, between the observationally derived albedo values for reflection nebulae and the values
predicted by Kim \& Martin. This may simply reflect the fact that their assumed
graphite/silicate composition for the grains does not correspond entirely to reality.  
Nevertheless,
we have for the first time a relatively clear indication that the dust albedo in the UV in the
diffuse ISM of the Galaxy is lower than the albedo in denser regions, about 0.45 vs.\ 0.65, and
that this difference is related to the typical grain sizes found in these environments, with 
smaller
grains in the diffuse ISM yielding a lower albedo.

     Additional support for these conclusions is found in the observations and analysis of scattered
far-UV radiation seen at high galactic latitudes by Onaka and Kodaira (\cite{OK91}).  While more
limited in spatial coverage, their data set allowed them to conclude $a~\geq~0.32$ and $g~\geq~0.5$,
in agreement with our results.  The suggestion of smaller dust grains, on average, at higher $|$z$|$-
distances also finds support in the finding by Kiszkurno \& Lequeux (\cite{KL87}) of a general
steepening of the far-UV extinction curve with increasing $|$z$|$-distance from the galactic plane.
This characteristic is attributed to a systematic shift toward smaller grain sizes by these authors.  

     When comparing the phase function asymmetry values in Table 3, no clear difference can be
established, and, in fact, may not exist.  Within existing uncertainties, we conclude that the phase
function asymmetry of scattering by dust in the diffuse ISM and in dense nebulae is identical.
The phase function asymmetry is determined solely by
those grains which contribute to the scattering.  If, for example, the diffuse ISM contains a
separate population of very small grains ($\ll \lambda$) which would provide absorption, the albedo
would be lowered while the phase function would still entirely be determined by the larger
scattering grains and would remain unchanged.  In addition,
a very  small change in $g$ represents a significant change in the phase function when
$g~\approx$~0.7, and as long as any determination of $g$ is uncertain to $\pm$~0.1, 
any real differences equal to small $\Delta g$'s will
not likely be detectable.  The only situation which allows more precise differences in $g$ to be
derived is the determination of relative $g$-values as a function of wavelength in a given object
(e.g. \cite{CBG95}).

\subsection{The DGL-N(HI) Correlation}

     The positive correlation between the far-UV background intensity and the column density
of neutral hydrogen is generally viewed as evidence for the Galactic origin of a major part of
the UV background.  (\cite{Leq90}; Bowyer 1991).  However, a number of separate
investigations (Bowyer 1991, Table 1) have produced a wide range of values for the slope of the
DGL-N(HI) correlation, often mutually exclusive.  Bowyer (\cite{B91}) tabulates values ranging from
0.3 to 1.8 [photon units/10$^{18}$ HI cm$^{-2}$] from eleven different studies.  To this, we now add our
present result of 2.5~$\pm$~0.5 [photon units/10$^{18}$ HI cm$^{-2}$], larger than any found previously.

     The significance of these apparent discrepancies is easily understood.  The scattered light
intensity per [10$^{18}$ HI cm$^{-2}$] is determined by the scattering cross section per H-atom, the 
incident
radiation field, and the phase function asymmetry $g$.  A relatively large value of $g$, as we have
deduced in this investigation, provides a strong coupling between the anisotropy of the
illuminating radiation field and the anisotropy of the scattered radiation field.  The FAUST
observations were conducted in a region of the UV sky where the gradient in illuminating
intensity between near-equatorial galactic latitudes and polar latitudes is one of the strongest 
in
the entire sky.  The fact that this strong gradient is indeed reflected in a larger value for 
the
slope in the DGL-N(HI) correlation is by itself an independent indication that the scattering
phase function must be relatively large.  Correlations between measured UV background
intensities with N(HI) are therefore comparable only, if the data have been obtained from
identical regions of the sky.

     We can demonstrate that our value for the slope of the DGL-N(HI) correlation is well within the
range expected for different regions of the sky with the following experiment.  For this test, 
we
adopted scattering properties $a$~=~0.45, $g$~=~0.6 and the average N(HI) values of the FAUST
fields.  We predicted the observable DGL, including 300 units of extragalactic background, for
the actual FAUST fields and found a slope for the DGL-N(HI) correlation of 2.58 
[photon units/10$^{18}$ HI cm$^{-2}$] (Fig. 10). 
\begin{figure}[htb]
\plotone{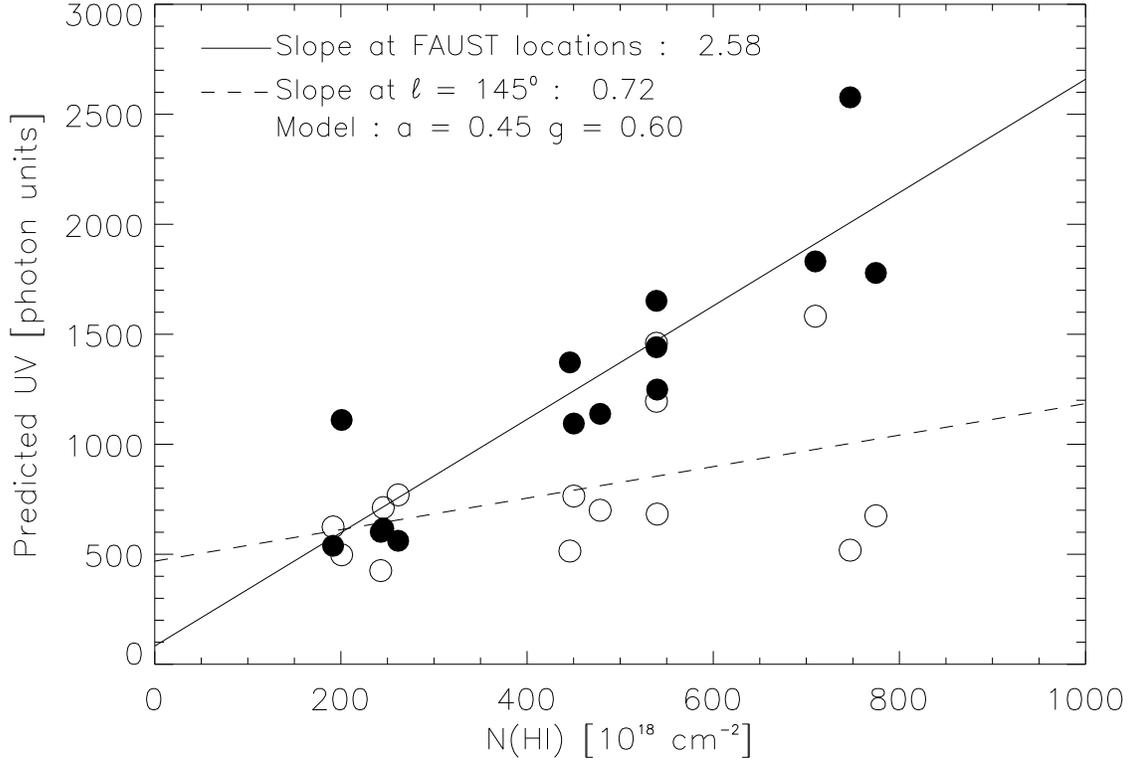}
\caption[f10.eps]{Predicted correlation between UV DGL and N(HI) for two different Galactic
regions, assuming $a$~=~0.45, $g$~=~0.60; the actual positions of the FAUST fields (filled circles;
slope~=~2.58~[photon units/10$^{18}$cm$^{-2}$]) and hypothetical positions with the Galactic latitudes and
N(HI) values identical to the FAUST positions, but moved to $\ell$~=~145$\arcdeg$ (open circles;
slope~=~0.72~[photon units/10$^{18}$cm$^{-2}$).   \label{fig10}}
\end{figure}
Retaining the galactic latitudes of the FAUST fields as well as their N(HI) values, we calculated
the DGL (plus 300 units extragalactic background) under the assumption that their galactic
longitude was $\ell~=~145\arcdeg$.  This is a region where the illuminating radiation field has a much
weaker gradient with latitude.  The resulting slope in the DGL-N(HI) correlation was now 
0.72 [photon units/10$^{18}$ HI cm$^{-2}$],
near the lower end of the range tabulated by Bowyer (\cite{B91}).  This clearly identifies the
anisotropy of the illuminating radiation field as the dominant cause of the anisotropy of the 
DGL
in the presence of a strongly forward-directed scattering phase function.  The high value of the
slope of the DGL/IRAS 100$\mu$m correlation found by Sasseen \& Deharveng (\cite{SD96}) for these
FAUST data is also consistent with the picture of anisotropic illumination coupled with a high
value of $g$.

\subsection{The Extragalactic Background}

     The present study did not permit a separate determination of the airglow and the partially
attenuated extragalactic background components, except that their sum should be about 700 photon units
averaged over the 14 FAUST fields.  The reason for this is two-fold:  We lacked a spectroscopic
capability which would have been able to separate the known line emissions of the airglow from
the quasi-continuous extragalactic background spectrum; and the variations in airglow intensity
among different FAUST fields were large in comparison with the weak Galactic latitude
dependence in the extragalactic background imposed by Galactic foreground extinction.  We
assumed an extragalactic component of 300 units in agreement with the determination by WP (\cite{WP94}) 
and the summary of earlier results by Paresce (\cite{Par90}).  This produced the
deduced airglow intensities listed in Table 2.  After correction for Galactic foreground 
extinction, 
the extragalactic component is reduced to about 230 units for the high-latitude fields and to 
about
100 units for the FAUST fields with the highest N(HI) values leading to an average for our sample of
162~$\pm$~46 photon units.  

     A discussion of the origin of the extragalactic component and the potential
contributors is beyond the scope of the present paper.  However, we must examine whether the 
assumption of an extragalactic component, anticorrelated with N(HI) through extinction, has a
significant effect upon the final scattering characteristics derived from the DGL analysis.
Repeating our analysis with the assumption that an extragalactic component in the UV is completely
absent, we find that the excess intensity must then be mostly attributed to a higher airglow.
The resultant albedo and phase function asymmetry are essentially unchanged, 0.43 instead of
0.45, and 0.65 instead of 0.675, respectively.  Thus, should future studies reveal our assumption 
of 300 units for the extragalactic UV background as inaccurate, this would not affect the validity
of the present results.

     On the other hand, an extragalactic component of {\it more} than 300 units is effectively excluded
by the lower limit to the total far-UV background, including extragalactic, measured by Onaka
and Kodaira (\cite{OK91}) who directly recorded intensities as low as 300 units at high galactic
latitudes, effectively using the same bandpass.  It is important to state, however, that it is not
necessary to know the intensities of the airglow and the extragalactic background separately in
order to determine the dust scattering properties of galactic dust, provided two conditions are met:
(1) the observations must cover a sufficiently large range of galactic latitudes to allow the dependence
of the measured intensity upon the HI column density to be revealed, and (2) the airglow component 
must be relatively constant during the observation.

\section{Conclusions}

     We have analyzed the 140-180nm UV background intensity in 14 FAUST fields at
intermediate and high Galactic latitudes after contributions from discrete sources such as 
stars
and galaxies had been removed.  We conclude the following:

     (1) We confirm that a major component of the measured UV background must be of
Galactic origin based on its strong correlation with N(HI), with cosec$|b|$, and with the IRAS
100$\mu$m background intensity.  We interpret this component as DGL, produced by scattering of
stellar photons by Galactic dust.

     (2) The slope in the FAUST DGL-N(HI) correlation of 2.5~$\pm$~0.5 [photon units/10$^{18}$~HI
~cm$^{-2}$] is found to be steeper than that seen in earlier investigations in other Galactic 
regions. 
We explain this discrepancy as owing to the existence of a very strong Galactic latitude 
gradient
in the illuminating radiation field in the FAUST region.

     (3) We found our radiative transfer model capable of matching the detailed variation of
the DGL intensity from one line-of-sight to another.  The most important characteristic of the
model responsible for this is the incorporation of the detailed anisotropy of the interstellar
radiation field in the UV and its variation with z-distance from the Galactic plane.

     (4) The radiative transfer analysis of the DGL component with our model resulted in a
set of well-constrained scattering parameters for dust in the diffuse ISM at intermediate and 
high
Galactic latitudes:  albedo $a~=~0.45~\pm~0.05$ and phase function asymmetry $g~=~0.68~\pm~0.10$.

     (5) The dust albedo is some 50\% lower than that commonly derived for dense reflection
nebulae and star forming regions.  We interpret this difference as arising from a difference in
size distributions with the grains in the diffuse ISM being smaller on average.

     (6) A contribution of about 700~$\pm$~200 photon units appears to be uncorrelated with
Galactic lines-of-sight.  We interpret this component  as due to a sum of residual airglow (530
~$\pm$~190 photon units) and isotropic extragalactic background radiation (160~$\pm$~50 photon units,
after correction for Galactic extinction).

We thank Jens Petersohn for his invaluable assistance with the WP radiative transfer model
during the early stages of this work.  The referee, Dr. Ken Sembach, provided a number of thoughtful 
and constructive comments which helped to improve the presentation of this paper, and for which
we are grateful.
ANW and BCF acknowledge material support from NASA LTSA Grant NAGW-3168
to The University of Toledo.

\end{document}